\documentclass{amsart}

\usepackage{amssymb}
\usepackage{amstext}
\usepackage{comment}

\newcommand{\paren}[1]{\left(#1\right)}
\newcommand{\biggparen}[1]{\bigg(#1\bigg)}

\newcommand{\bra}[1]{\left<#1\right|}
\newcommand{\ket}[1]{\left|#1\right>}
\newcommand{\braket}[2]{\left<#1|#2\right>}
\newcommand{\ketbra}[2]{\left|#2\right>\!\!\left<#1\right|}

\renewcommand{\choose}[2]{\left(\begin{matrix}#1\\#2\end{matrix}\right)}

\newcommand{\B}{\mathbb{B}}
\newcommand{\C}{\mathbb{C}}
\newcommand{\N}{\mathbb{N}}
\newcommand{\E}{\mathbb{E}}

\newcommand{\Z}{\mathbb{Z}}

\newcommand{\basis}{\B^{\Z}}
\newcommand{\newfield}{\N\to\C}

\newcommand{\truespace}{\mathcal{H}}
\newcommand{\scalespace}{\mathcal{S}}
\newcommand{\scalespaceops}{\mathcal{T}}
\newcommand{\mpsspace}{\mathcal{M}}
\newcommand{\mpsops}{\mathcal{O}}
\newcommand{\subring}{\mathcal{R}}
\newcommand{\linearof}{\mathcal{L}}

\newcommand{\vb}{b}
\newcommand{\vO}{O}
\newcommand{\vs}{s}
\newcommand{\vt}{t}
\newcommand{\vz}{z}

\newcommand{\trace}{\text{tr}}

\begin{document}

\title{Chasing infinity with matrix product states by embracing divergences}
\author{Gregory M. Crosswhite}

\begin{abstract}
In this paper, we present a formalism for representing infinite systems in quantum mechanics by employing a strategy that embraces divergences rather than avoiding them.  We do this by representing physical quantities such as inner products, expectations, etc., as \emph{maps} from natural numbers to complex numbers which contain information about \emph{how} these quantities diverge, and in particular whether they scale linearly, quadratically, exponentially, etc. with the size of the system.  We build our formalism on a variant of matrix product states, as this class of states has a structure that naturally provides a way to obtain the scaling function.  We show that the states in our formalism form a module over the ring of functions that are made up of sums of exponentials times polynomials and delta functions.  We analyze properties of this formalism and show how it works for selected systems.  Finally, we discuss how our formalism relates to other work.
\end{abstract}

\maketitle

\tableofcontents

\section*{Introduction}

Infinitely large systems in quantum mechanics are unlikely to ever exist in a laboratory, but they are nonetheless very important from a theoretical perspective because they can provide useful information about the scaling properties of their finite counterparts, as well as a way to study the bulk properties of a system independently from finite-size effects.  There are many formalisms for representing an infinite system such as von Neumann tensor product spaces \cite{Neumann1939} (which refuse to admit any states whose norms do not converge), finitely correlated states \cite{Fannes1992} \cite{Fannes1994} (which represent states as functionals that specify expectations of local operators), and infinite matrix product states \cite{Vidal2007} \cite{Orus2008} \cite{Crosswhite2008} \cite{McCulloch2008} (which are essentially a variant of finitely correlated states where the boundaries are a function of the repeated tensor, which both makes them useful as a simulation ansatz and allows for expectations to be computed for arbitrary matrix product operators).

These formalisms, however, are in a sense limited in the states that they can represent because of the need to keep quantities of interest convergent despite the infinite size of the system.  In this paper we introduce a formalism with a different philosophy:  rather than seeking to eliminate the divergences, we \emph{embrace} them as an essential part of the formalism.  We do this by having our space of coefficients be the ring of maps from natural numbers to complex numbers (i.e. $\N\to\C$) rather than the field of complex numbers.  This works out because the information that we usually want from a state is exactly how quantities of interest diverge with respect to the system size.

The remainder of this paper is organized as follows.  In Section \ref{sec:motivation} we motivate the key idea of this formalism --- namely, changing the space of coefficients from $\C$ to $\N\to\C$.  In Section \ref{sec:formalism} we present the formalism itself.  In Section \ref{sec:properties} we describe some of the properties of this formalism.  In Section \ref{sec:examples} we show that the formalism gives the expected results for some example states and operators.  In Section \ref{sec:other-work} we compare this formalism to similar work that has been done by others.

\section{Motivation}
\label{sec:motivation}

A fundamental problem in studying infinite systems is that quantities of interest will in general diverge.  To explore this in a more concrete setting, let us consider an infinitely large 1D system of spin-$\frac{1}{2}$ particles.  Since the state of each particle lives in the space $\C^2\equiv\text{span}\{\ket{0},\ket{1}\}$, the state of the whole system lives in the informal\footnote{By ``informal'' here we mean that at this time this space only exists in a rough conceptual sense, rather than a formal sense, in order to provide a setting for an informal discussion.  In the next section we will take the ideas discussed in this section and give them formal grounding.} infinite tensor product space $\truespace=\left(\C^2\right)^{\otimes\,\Z}\,\,$\footnote{This is shorthand for $\bigotimes_{\Z}\C^2$, which is the tensor product of an infinite number of copies of $\C^2$;  the presence of $\Z$ instead of $\N$ indicates that the system extends infinitely both to the left and to the right.}.   This space is not immediately a Hilbert space because the obvious choice of inner product,
$$\paren{\,\,\bigotimes_{k\in\Z}\bra{\psi_k}}\paren{\,\,\bigotimes_{k\in\Z}\ket{\phi_k}} := \prod_{k\in\Z} \braket{\psi_k}{\phi_k},$$
involves an infinite number of factors and so in general it will be divergent;  for example, if we set $\ket{\phi_k} := 2\ket{0}$ then we have that,
$$\braket{\phi}{\phi}=\biggparen{2\bra{0}}^{\otimes\,\Z}\biggparen{2\ket{0}}^{\otimes\,\Z} = \prod_{k\in\Z} 4 = \infty.$$

One rather drastic solution to fix this problem is to (roughly speaking) simply remove everything from our set of states that could possibly result in a divergent inner product, resulting in a new space $\bar{\truespace}$.  This is the approach taken by von Neumann \cite{Neumann1939}, who proved that this procedure results in a Hilbert space.  To illustrate what states in this space can look like, a non-trivial example of a state in $\bar{\truespace}$ is
$\bigotimes_{k\in\Z}\ket{\psi_k},$ where $$\ket{\psi_k} := \paren{1+\frac{i}{2^{|k|/2}}}\ket{0}$$
(with $i$ denoting the imaginary number, not an index variable).  To show that this is a well-defined member of $\bar{\truespace}$, it suffices to show that the norm,
$$\left\|\bigotimes_{k\in\Z}\ket{\psi_k}\right\|=\prod_{k\in\Z}\left|1+\frac{i}{2^{|k|/2}}\right|,$$
is convergent.  To do this, we use the result of Lemma 2.4 (p.13) of \cite{Neumann1939}, that an infinite product converges when each factor is greater than 1 and the sum over each factor minus 1 is convergent, and then we apply the triangle equality to see that, $$\sum_{k\in\Z}\paren{\left|1+\frac{i}{2^{|k|/2}}\right|-1}\le\sum_{k\in\Z} \frac{1}{2^{|k|/2}},$$ which is a convergent series.

Unfortunately, there are many aspects of this solution that are unsatisfying.  First, it excludes many states that are important from a theoretical (quantum information) perspective, but which have a divergent norm such as the so-called W-state,
$$\ket{W} := \sum_{j\in\Z}\bigotimes_{k\in\Z}\begin{cases}\ket{1} & j= k \\ \ket{0} & \text{otherwise}.\end{cases}$$

Second, although normalizations of all states in $\bar{\truespace}$ are convergent, \emph{expectations} with respect to physically relevant observables on $\bar{\truespace}$ are not convergent in general.  For example, consider a system in a magnetic field which has the Hamiltonian $H := \sum_{k\in\Z} \sigma_{Z}^k,$ where $\sigma_{Z}^k$ denotes the Pauli $Z$ spin matrix acting on the $k$-th site in the lattice.  For even the simplest state, $\ket{0}^{\otimes\,\Z}$, the expected energy is infinite.  Fortunately, in this case there is an easy answer:  change the question one is asking to the expected energy \emph{per site}, to which one gets a finite and meaningful answer (namely, $+1$).  However, consider a system with long-range spin-coupling interactions along the $Z$ axis given by $H := \sum_j^{\Z}\sum_{k\ne j}^{\Z} \sigma_Z^j \sigma_Z^k$; in this case, even the energy per site diverges for the state $\ket{0}^{\otimes\,\Z}\,\,$\footnote{We left off a factor that weakens the interaction with distance in order to make the analysis simpler when we return to this hamiltonian later, but it is worth observing that one also obtains a divergence if the strength of the interaction term falls off like the reciprocal of the distance between the particles, since $\sum_{r=1}^\infty \frac{1}{r}=\infty$.}.

Finally, $\bar{\truespace}$ has the inconvenient property that given a non-trivial state $\Psi(\alpha):=\paren{\alpha\ket{0}}^{\otimes\,\Z}\in\bar{\truespace}$ with $\alpha\in\mathbb{R^+}$, we are restricted to $\alpha=1$, since if $\alpha>1$ then $\|\Psi(\alpha)\|$ is divergent, and so $\Psi(\alpha) \notin \bar{\truespace}$, and if $\alpha<1$ then $\|\Psi(\alpha)\| = 0$, and so $\Psi(\alpha)=0$ is trivial.  This is unfortunate because it implies that in general
$$\left\|\frac{\Psi(\alpha)}{\sqrt{\Psi(\alpha)^\dagger\Psi(\alpha)}}\right\|\neq 1,$$
which means that, unlike the case of a finite tensor product space, we cannot disregard the normalization of the (infinite tensor product) factors composing a state during a computation under the assumption that we can take care of it at the end by dividing out the overall normalization.

All of the problems described above are essentially an inescapable consequence of the attempt to design our state space in such a way that the inner products between states and the expectation values of observables do not diverge.  However, given that we are considering an imaginary infinite physical system, there really is no reason for these quantities \emph{not} to diverge.  In fact, many of the quantities that we care about \emph{will} diverge in general, and this is perfectly acceptable because what we really care about is not \emph{whether} a given quantity diverges, but \emph{how quickly} it diverges as a function of the size of the system.  For example, is the divergence linear or quadratic?  What is the coefficient on the highest order term?  And so on.  We thus see that the solution is not to eliminate the divergences in the state space but to \emph{embrace} them by changing the underlying set of vector coefficients from the field of complex numbers to the commutative ring of maps from natural numbers to complex numbers, $\newfield$, with point-wise addition and multiplication --- i.e., $(f+g)(n) := f(n)+g(n)$, $(fg)(n) := f(n)g(n)$, the additive identity is $n\mapsto 0\,\,$\footnote{The notation $n\mapsto\cdot$ refers to the map that takes $n$ to $\cdot$, used when we don't want to give a function a name;  for any function $f\in\newfield$ we have that $f\equiv n\mapsto f(n)$.}, and the multiplicative identity is $n\mapsto 1$.  Note that this is not a field because, for example, $n\mapsto\delta_{n0}$ and $n\mapsto\delta_{n1}$ are non-zero elements, but their quotient, $[n\mapsto\delta_{n0}]/[n\mapsto\delta_{n1}]=n\mapsto\delta_{n0}/\delta_{n1}$, is undefined for $n\neq1$.  Nonetheless, there is a subset of elements which can be divided:  given $f,g\in\newfield$, the quotient $$(f/g)(n):=n\mapsto\begin{cases} f(n)/g(n) & g(n)\ne 0 \\ 0 & g(n) = 0,\end{cases}$$ is well defined if and only if $g(n)=0\Rightarrow f(n)=0$.

Making this change to $\truespace$, however, is not quite enough, because although it gives us the correct \emph{type} for the inner product, it does not give us the \emph{value} of the inner product between two arbitrarily chosen states.  We shall present a formal construction in the next section, but first, to motivate how the inner product \emph{should} work, it is worthwhile to revisit the problems discussed in the previous section.

First we revisit the W-state.  For any finite system of size $n$, the norm of the unnormalized W-state is $n$ because there are $n$ orthogonal terms, i.e.,
$$\left\|\sum_{j=1}^n \bigotimes_{k=1}^n \begin{cases} \ket{1} & j=k \\ \ket{0} & \text{otherwise} \end{cases} \right\| = n.$$  Thus, the value for the norm of the infinite W-state should be $n\mapsto n$.

Second we revisit observables.  We start by considering the expectation of the external magnetic field hamiltonian, $H := \sum_{k\in\Z} \sigma_{Z}^k$, with respect to $\ket{0}^{\otimes\,\Z}$.  Using the same reasoning as for the W-state, we see that since for any finite system there are $n$ terms each with an expectation value of $+1$, we conclude that the expectation of the infinite system should be $n\mapsto n$.

Next we consider the expectation of the spin-coupling hamiltonian, $$H := \sum_{j\in\Z}\sum_{k\ne j} \sigma_Z^j \sigma_Z^k,$$ with respect to $\ket{0}^{\otimes\,\Z}$.  Since for any finite system there are $n(n-1)$ terms in the hamiltonian, each of which has expectation $+1$, we conclude that the expectation of the infinite system should be $n\mapsto n^2 - n$.

Now we consider the expectation of the external magnetic field hamiltonian, but with respect to the unnormalized W-state.  First, we define the following shorthand notation,
$$\gamma_{ij}^{a,b} = \begin{cases}
b & i=j \\
a & \text{otherwise}.
\end{cases}$$  For a finite system of size $n$ we have that,
\begin{equation*}
\begin{aligned}
\left<W\right|H\left|W\right>
&= \sum_{j=1}^n\bigotimes_{k=1}^n\bra{\gamma_{jk}^{0,1}}\cdot
\sum_{l=1}^n \sigma_{Z}^l\cdot
\sum_{m=1}^n\bigotimes_{o=1}^n\ket{\gamma_{mo}^{0,1}}, \\
&=
\sum_{l=1}^n \gamma_{lm}^{+1,-1}\cdot
\sum_{j=1}^n\bigotimes_{k=1}^n\bra{\gamma_{jk}^{0,1}}\cdot
\sum_{m=1}^n\bigotimes_{o=1}
\ket{\gamma_{mo}^{0,1}}, \\
&= \sum_{l=1}^n \gamma_{lm}^{+1,-1}\cdot
\sum_{j,m=1}^n\delta_{jm}, \\
&= \sum_{l,m=1}^n \gamma_{lm}^{+1,-1} = n(n-2),
\end{aligned}
\end{equation*}
and so the unnormalized expectation value of the infinite system should be $n\mapsto n(n-2)$.  Since the norm of the W state is $n\mapsto n$, the (normalized) expectation value is given by
$$\frac{\left<W\right|H\left|W\right>}{\left<W|W\right>} = \frac{n\mapsto n(n-2)}{n\mapsto n} = n\mapsto n-2.$$  (Recall that this quotient is well-defined because the numerator is zero whenever the denominator is zero.)

Finally, we revisit unnormalized states.  We now have a way of capturing explicitly the norm of $\Psi(\alpha):=(\alpha\ket{0})^{\otimes\,\Z}$.  For any finite system we have that $\|\alpha\ket{0}^{\otimes\,n}\|=\alpha^n$, and so $\|\Psi(\alpha)\|=n\mapsto\alpha^n$.  In particular, observe that now we have that
$$
\left\|\frac{\Psi(\alpha)}{\sqrt{\Psi(\alpha)^\dagger\Psi(\alpha)}}\right\|
=\frac{(n\mapsto \alpha^n)}{\sqrt{(n\mapsto \alpha^n)(n\mapsto\alpha^n)}}
= n\mapsto 1,$$ where the square-root is understood to operate pointwise on its argument (i.e., $\sqrt{n\mapsto f(n)} := n\mapsto\sqrt{f(n)}$).

\section{Formalism}
\label{sec:formalism}

Up until now we have been working in an informal setting in order to get an intuitive feel for how our state space should work;  in this section we shall formally define our space in such a way as to agree with our intuition.

First we modify our space $\truespace$ to explicitly have coefficients in the commutative ring $\newfield$ by redefining it as a module\footnote{If you are not familiar with the theory of modules, you can just think of them as a generalization of vector spaces where the set of coefficients is a ring rather than a field --- that is, where the set of coefficients may lack the division operation.} over $\newfield$ given by $\truespace := [\newfield]^{\basis\times\Z}$ where $\B := \{0,1\}\,\,$\footnote{The choice of $\B$ was made for concreteness;  for the general case (qudits) one can obviously use a larger basis set.} (i.e., $\B$ is shorthand for the basis of a qu$\B$it);  the notation $A^B$ is used to equivalently represent the direct product\footnote{Note that the direct product operation, denoted by $\prod$, is not the same as the tensor product operation, denoted by $\bigotimes$;  the difference is that in a direct product space, addition is a pointwise operation, i.e. $(a_1,b_1)+(a_2,b_2)=(a_1+a_2,b_1+b_2)$, whereas in a tensor product space in general you cannot combine a sum of elements into a single element unless they share common elements, i.e. $(a_1,b) + (a_2,b)=(a_1+a_2,b)$.} $\prod_B A$ and the set of maps $B\to A$.  The way to think of elements in $\truespace$ is that they assign coefficients for each basis vector that depend on what part of that vector you are looking at, so to obtain a particular coefficient you start with some basis vector $\vz\in\basis$ (for example, the vector that is 1 at position 5 and 0 everywhere else) whose coefficients you are interested in learning about, and then you pick a starting location in this vector, $i\in\Z$;  this then gives you an element of the ring $\newfield$, which for each size $n$ gives you a complex value corresponding to the weight of $\vz_{i:i+n}\in\B^{\otimes\,n}$, which denotes the slice of of the basis vector, $\vz$, that starts at position $i$ and ends at position $i+n-1$.  It is straightforward to show that this map is linear (with respect to elements in $\truespace$).

$\truespace$ has the right type of coefficients, but it is missing the structure needed to give it a well-defined inner product.  To get this structure, we construct a new space that consists of a direct product of a sequence of increasingly large systems, $\scalespace := \prod_{n\in\N} \scalespace_n$ where $\scalespace_n := (\C^2)^{\otimes\,n}\,\,$\footnote{This might look a lot like a Fock space, but it is constructed using a direct product operation whereas Fock spaces are constructed using a direct sum operation.  The difference between these operations is that the states in a direct sum can only be non-zero in a finite subset of the spaces being summed, whereas a direct product has such restriction;  it follows directly that for finite-dimensional spaces the direct sum and the direct product are the same.}.  We will henceforth make implicit use of the natural injections $\B\to\C^2$ given by $b\mapsto\ket{b}$ and $\B^{\otimes\,n}\to(\C^2)^{\otimes\,n}$ given by $\vb\mapsto\bigotimes_{i=1}^n\ket{\vb_i}$.  $\scalespace$ is a module over the ring $\newfield$ where addition acts point-wise (i.e., $(\vs+\vt)_n :=\vs_n+\vt_n$) and multiplication by elements in $\newfield$ is given by $(f\cdot\vs)_n :=f(n)\vs_n$;  it is straightforward to show that the module laws hold.  We define an inner-product operation on $\scalespace_n$ as follows: given elements $\vs,\vt\in\scalespace$, the inner-product is given by $\paren{\vs\cdot\vt}_n:=\vs_n\cdot\vt_n$, and it is straightforward to show that this operation is multilinear.  While we are at it, we shall also define a subset $\scalespaceops\subset\linearof(\scalespace)$, where $\linearof(\scalespace)$ denotes the set of all linear operators acting on $\scalespace$, as the set $\scalespaceops:=\prod_{n\in N}\scalespaceops_n$ where $\scalespaceops_n:=\linearof(\scalespace_n)$ and given $\vO\in\scalespaceops$ and $\vs\in\scalespace$ we have that $[\vO(\vs)]_n := \vO_n(\vs_n)$.\footnote{Note that not all operators in $\linearof(\scalespace)$ take this form;  for example, the operator $O:=\cdot\otimes\ket{0}$, i.e. $[O(\vs)]_0 :=0$ and $[O(\vs)]_n := \vs_{n-1}\otimes\ket{0}$ for $n>0$.}

We now relate elements in $\scalespace$ back to $\truespace$ via a map $F:\scalespace\to\truespace$ which is given by $F(\vs) := (\vz,i)\mapsto [n \mapsto\vs_n\cdot\vz_{i:i+n}]$. To understand what is meant by this, recall that because $\truespace$ is defined in terms of a direct product, every element in it is equivalent to a map, $\basis\times\Z\to[\newfield]$, and so we specify the value of $F(\vs)$ by defining a map from each pair $(\vz,i)\in\basis\times\N$ to a value in $\newfield$, which itself is a map that, for every $n\in\N$, is equal to the overlap of $\vs_n$ and the slice of $\vz$ of length $n$ starting at $i$.  In short, for each element $\vs\in\scalespace$, we map it to an element of $\truespace$ such that, for every infinite basis vector $\vz\in\B^\Z$ and every starting position $i\in\Z$, the weight for each $n$ is the overlap between $s_n$ and $\vz_{i:i+n}$.

We are now closer to what we want, but there is no additional structure in $\scalespace$ that makes it correspond to a sequence of systems converging in some sense to the infinite system;  for example, a valid member $\vs\in\scalespace$ is given by
$$\vs_n=\begin{cases}\ket{0}^{\otimes\,n}&n\,\,\text{even}\\ \ket{1}^{\otimes\,n}&n\,\,\text{odd},\end{cases}$$ but clearly there is no infinite state that corresponds even loosely to the notion of an infinite limit of this sequence.  Thus, we shall instead work in a subspace $\mpsspace\subset\truespace$ where all elements in the sequence essentially follow the same pattern;  this space is a form of infinite matrix product state\footnote{In this paper we will present our variant of matrix product states starting from the ground up so that no prior knowledge is required, but if the reader is interested for more information on matrix product states and algorithms, then see \cite{Crosswhite2007}, \cite{Schollwock2005}, \cite{White1992}, \cite{Perez-Garcia2007}, and \cite{Verstraete2004}.}, and comparisons between the particular construction we are about to present and other work is provided in Section \ref{sec:other-work}.

Our formalism will be constructed in the following steps.  First, we shall define the form that states take in $\mpsspace$, show how they relate back to states in $\scalespace$ via a map $G:\mpsspace\to\scalespace$, and then define a subring $\subring\subset\newfield$ such that $\mpsspace$ is a module over $\subring$ and the map $G$ is linear.  Second, we shall introduce another characterization of $\subring$ that relates it to the set of exponential polynomials.  Third, we shall define an inner-product operation over arbitrary elements of $\mpsspace$ and show that is consistent with the inner product in $\scalespace$.  Finally, we shall define operators over $\mpsspace$ (which automatically gives us density matrices).

\subsection{Definition of states}

A state $\ket{\psi}\in\mpsspace$ is defined as a tuple $\psi := (m,L,M,R)$, where $m\in\N$, $L,R\in\C^m$, and $M\in \B\to \C^{m\times m}\,\,$\footnote{The presence of both $L$ and $R$ actually turns out to be redundant because, as long as $L\ne 0$ or $R\ne 0$, there always exists a transformation that results in an equivalent state with either $L_i=\delta_{i1}$ (if $L\ne 0$) or $R_i=\delta_{i1}$ (if $R\ne 0$).  Specifically, one can construct an invertible matrix $X$ with the property that $(L\cdot X)=\delta_i$ (by putting $L$ in the first column, dividing by $|L|^2$, and then filling the rest of the matrix with an orthonormal basis orthogonal to the first column) such that the state $\tilde\psi =(\tilde m,\tilde L,\tilde M,\tilde R)$ given by $\tilde m:= m$, $\tilde L = L\cdot X$, $\tilde R := X^{-1} \cdot R$, $\tilde M(i,j) := X^{-1}\cdot M(i,j)\cdot X$ has the property that $G(\tilde\psi)=G(\psi)$ and $\tilde L_i =\delta_{i1}$, and one can do likewise for $R$ instead.  We will nonetheless keep both $L$ and $R$ around because it keeps the rule for adding states simple, as otherwise we would have to perform the described canonicalization transformation after every sum.}.  States in $\mpsspace$ can be mapped to $\scalespace$ via the function $G$, given by
$$\big[G(\ket{\psi})_n\big]_{i_1\dots i_n} := L\cdot M(i_1)\cdot M(i_2)\cdots M(i_n)\cdot R.\,\, \footnote{We here borrow the convention of \"Oslund and Rommer in \cite{PhysRevLett.75.3537} and \cite{Rommer1997} of using function notation to capture the `third' dimension of the tensor $M$.}$$

Now, define the ring $\subring\subset\N\to\C$ to be the set of all functions such that for every function $f\in\subring$ there exists a tuple $(m,L,M,R)$, where $m\in\N$, $L,R\in\C^m$, and $M\in\C^{m\times m}$ such that
$$f(n) = L\cdot M^n\cdot R.$$

For any $f,g\in \subring$, let the sum $h:=f+g$ be defined by $m_h=m_f+m_g$, $L_h := L_f\oplus L_g$, $M_h(i) := M_f(i)\oplus M_g(i)$, and $R_h := R_f\oplus R_g$, and let the product $f\cdot g$ be defined like the sum but with $m_h:=m_f\cdot m_g$ and the direct sums replaced with tensor products.  With this structure $\subring$ is a commutative ring (with the proof of the laws left as an exercise for the reader), and furthermore, since $(f+g)(n)=f(n)+g(n)$ and $(f\cdot g)(n)=f(n)g(n)$ we have that $\subring$ is not only a subset, but a sub\emph{ring} of $\N\to\C$.

Next we note that $\mpsspace$ is an abelian group when endowed with addition such that for all $\ket{a}, \ket{b}\in\mpsspace$ we have that $\ket{c} := \ket{a} + \ket{b}$ where $m_c=m_a+m_b$, $L_c=L_a\oplus L_b$, $M_c(i)=M_a(i)\oplus M_b(i)$, and $R_c=R_a\oplus R_b$;  it is straightforward to show that for all $n\in\N$ we have that $G(\ket{a}+\ket{b})_n = G(\ket{a})_n + G(\ket{b})_n$.

Finally, we define left-multiplication on $\mpsspace$ such that given $c\in \subring$ and $\ket{x}\in\mpsspace$ we have that $\ket{y}:=c\ket{x}$ is given by $m_y = m_c \cdot m_x$, $L_y = L_c \otimes L_x$, $M_y(z) = M_c \otimes M_x(z)$, and $R_y = R_c \otimes R_x$;  it is straightforward to show that, given these operations, $\mpsspace$ is a module over $\subring$ and that $G(c\ket{x})_n = c(n) G(\ket{x})_n$ and therefore $G$ is a linear map.

\subsection{Relation of $\subring$ to exponential polynomials}

In this subsection we shall introduce an alternative characterization of $\subring$.  Specifically, $\subring$ can equivalently be defined as the set of functions which take the form
\begin{equation}
\label{expolyform}
n\mapsto \sum_{k\in K}\lambda_k^n \cdot \text{poly}_k(n) + \sum_{l\in L}c_l \delta_{nl}
\end{equation}
where $\lambda_k\in\C$, $\text{poly}_k$ is a polynomial function, and $\delta$ is the Dirac delta function;  that is to say, $\subring$ is equivalent to the set of functions which can be expressed as a sum of exponential functions multiplied by polynomial functions and then added to a sum of delta functions.

This alternative characterization follows from the fact that all matrices are similar to a matrix in Jordan Normal Form, i.e. a matrix that is a direct sum of Jordan blocks.  It follows from this that for every matrix $M\in\C^{m\times m}$ there exists a unitary matrix $U$, an index set $K$, integers $m_k\in\N$ such that $\sum_{k\in K} m_k=m$, scalar values $\lambda_k\in\C$, and Jordan blocks $J_k\in\C^{m_k\times m_k}$ such that
$$M^n = U\cdot \paren{\bigoplus_{k\in K} J_k^n}\cdot U^\dagger$$
where
\begin{equation}
\label{jordanfunction}
\paren{J_k^n}_{ij} =
\begin{cases}
\choose{n}{j-i} \lambda_k^{n-(j-i)} & j \ge i \\
0 & j < i.
\end{cases}
\end{equation}
(See pages 385 and 386 in \cite{Horn1991} for a proof and explanation of the above, setting $p(\lambda):=\lambda^n$.) Furthermore, the unitary matrices and Jordan blocks can be computed numerically (though the computation is unstable when eigenvalues are very close to each other). Note this implies that if $\lambda_k \ne 0$, then $J_k^n$ is a matrix of polynomials of $n$, times exponentials of $n$ with base $\lambda_k$, and if $\lambda_k=0$, then $\lambda_k^{n-(j-i)}=\delta_{n(j-i)}$, and so $J_k^n$ is a matrix of delta functions.  Since the effect of multiplying by $L\cdot U$ on the left and $U^\dagger\cdot R$ on the right is to take a linear combination of elements from these matrices, we have therefore shown that if $f$ is in $\subring$ then $f$ takes the form of \eqref{expolyform}.

We now need to show the opposite direction, i.e. that every function that takes the form \eqref{expolyform} is in $\subring$.  To do this, it is sufficient to show how functions of the form $\lambda^n\cdot n^l$ and $\delta_{nl}$ are constructed, as a linear combination can then be taken to obtain any function in the form of \eqref{expolyform}.  We shall start by showing how to construct functions of the first form by induction.  For the base case, we observe that the constant function is easily constructed by letting $m=1$, $L_{1}=1$, $M_{11}=1$, and $R_1=1$.  For the inductive case, suppose we know how to construct $\lambda^n\cdot n^k$ for all $k<l$.  Let $m=l+1$, $L_i=\delta_{i1}\cdot l!\cdot\lambda^l$, $R_i=\delta_{i(l+1)}$, and $M$ be the $(l+1)\times (l+1)$ Jordan block with $\lambda$ in the diagonal.  By \eqref{jordanfunction}, we know that $L\cdot M^n\cdot R = \paren{\prod_{i=0}^{l-1} (n-i)}\lambda^n$;  this is a polynomial of degree $l$ with no coefficient on the leading-order term, and by the inductive hypothesis we can know that we can construct polynomials equal to all of the lower-order terms that we can subtract to cancel out all but the order $l$ term, and so we are done.

Next, to construct functions of the second form, $\delta_{nl}$, we let $m=l+1$, $L_i=\delta_{i1}{\choose{n}{l}}^{-1}$, $R_i=\delta_{i(l+1)}$, and $M$ be the $(l+1)\times(l+1)$ Jordan block with $\lambda=0$ in the diagonal;  then we have that
$L\cdot M^n\cdot R = \lambda^{n-l} =\delta_{nl}$.

\subsection{Definition of inner product}

Given two arbitrary elements $\ket{x},\ket{y}\in\mpsspace$, the inner product of $\ket{x}$ and $\ket{y}$ is a member of $\subring$ given by
\begin{equation}
\label{inner-product}
\braket{x}{y} = n\mapsto\paren{L_x^*\otimes L_y}\cdot\paren{\sum_{i\in\B} M_x^*(i)\otimes M_y(i)}^n\cdot\paren{R_x^*\otimes R_y},
\end{equation}
To see how \eqref{inner-product} is in $\subring$, observe that if we let $L:=L_x^*\otimes L_y$, $M:=\paren{\sum_{i\in\B} M_x^*(i)\otimes M_y(i)}$, and $R:=R_x^*\otimes R_y$, then \eqref{inner-product} can be rewritten in the form $n\mapsto L \cdot M^n\cdot R$, which is manifestly a member of $\subring$.

It is straightforward to show that the inner product defined above satisfies $\braket{x}{y}=\braket{y}{x}^*$; the property $\braket{x+y}{z}=\braket{x}{z}+\braket{y}{z}$ follows from the fact that tensor products are distributive over direct sums, and the property $\braket{\alpha x}{y}=\alpha\braket{x}{y}$ follows from the fact that the tensor product operation is associative.

Finally, observe that,
\begin{equation}
\label{eq:proof-operation-lifting}
\begin{aligned}
&G(\bra{x})\cdot G(\ket{y}) =\\
&\quad n\mapsto\sum_{i_1\dots i_n\in\B}\sum_{j_0\dots j_n=1}^{m_x}\sum_{k_0\dots k_n=1}^{m_y}\\&\qquad \left[(L_x^*)_{j_0}\cdot (M_x^*(i_1))_{j_0,j_1}\cdots(M_x^*(i_n))_{j_{n-1},j_n} (R_x^*)_{j_n}\right]\\
&\qquad\quad \cdot \left[(L_y)_{k_0}\cdot (M_y(i_1))_{k_0,k_1}\cdots(M_y(i_n))_{k_{n-1},k_n} (R_y)_{k_n}\right],\\
&\quad n\mapsto \sum_{i_1\dots i_n\in\B}\sum_{j_0\dots j_n=1}^{m_x}\sum_{k_0\dots k_n=1}^{m_y}\\&\qquad [(L_x^*)_{j_0}(L_y)_{k_0}]\cdot \left[\sum_{i_1\in\B} (M_x^*(i_1))_{j_0,j_1}(M_x^*(i_1))_{k_0,k_1}\right]\cdots\\
&\qquad\cdots \left[\sum_{i_n\in\B} (M_x^*(i_n))_{j_{n-1},j_n}(M_x^*(i_n))_{k_{n-1},k_n}\right]\cdot[(R_x^*)_{j_n}(R_y)_{k_n}],\\
&\quad n\mapsto\paren{L_x^*\otimes L_y}\cdot\paren{\sum_{i\in\B} M_x^*(i)\otimes M_y(i)}^n\cdot\paren{R_x^*\otimes R_y},\\
&\qquad = \braket{x}{y},
\end{aligned}
\end{equation}
so that the inner product on $\mpsspace$ is consistent with the inner product on $\scalespace$.

\subsection{Operators}

Rather than considering the set of all linear operators on the space $\mpsspace$, we shall instead focus our attention on the set of operators, which we shall call $\mpsops$, such that there exists a tuple $O:=(m_O,L_O,M_O,R_O)\in\mpsops$ where $m_O\in\N$, $L_O,R_O\in\C^{m_O}$, and $M_O\in\B\times\B\to\C^{m_O\times m_O}\,\,$\footnote{For more information on so-called matrix product operators see \cite{Crosswhite2007} and \cite{Murg2008}.}.  In an abuse of notation, we shall let $G$ also serve as a map from $\mpsspace$ to $\scalespace$ given by
$$\big[G(\ket{\psi})_n\big]_{i_1\dots i_n,\,\,j_1\dots j_n} := L\cdot M(i_1,j_1)\cdot M(i_2,j_2)\cdots M(i_n,j_n)\cdot R.$$

The operation of an operator $O\in\mpsops$ on a state $\ket{x}\in\mpsspace$ is given by $\ket{y}=O\ket{x}\in\mpsspace$ where $m_y=m_O\cdot m_x$, $L_y = L_O\otimes L_x$, $R_y =R_O\otimes R_x$ and $M_y = i\mapsto \sum_{j\in\B}M_O(i,j)\otimes M_x(j)\,\,$\footnote{It is worth observing that not all linear operators on $\mpsspace$ need take this form, such as the reversal operator that swaps $L$ and $R$ and transposes $M$ in the state's tuple.}. As with states, $\mpsops$ is a module over the ring $\subring$ using essentially the same constructions for sums and products with elements in $\subring$ as used for states;  additionally, given two operators, $O_1, O_2\in\mpsops$, we have that the product, $O:=O_1\cdot O_2=O_1O_2$, is given by $m_O = m_{O_1} \cdot m_{O_2}$, $L_O = L_{O_1}\otimes L_{O_2}$, $R_O =R_{O_1}\otimes R_{O_2}$, and $M_O = (i,j)\mapsto \sum_{k\in\B}M_{O_1}(i,k)\otimes M_{O_2}(k,j)$.  It is straightforward to show that $G$ is linear and furthermore that $G(O\ket{x})=G(O)G(\ket{x})$ and $G(O_1O_2)=G(O_1)G(O_2)$, where the latter equations can be proved by explicitly writing down the sums and regrouping the terms analogous to that done in Eq.  \ref{eq:proof-operation-lifting}.

Expectations follow straightforwardly from the rule for operating on a state and the rule for taking an inner product; specifically we have that $E:=\bra{\psi}O\ket{\psi}$ is given by $L_E = L_\psi^*\otimes L_O\otimes L_\psi$, $R_E = R_\psi^*\otimes R_O\otimes R_\psi,$ and
$$M_E = \sum_{i\in\B}\sum_{j\in\B} M_\psi^*(i)\otimes M_O(i,j)\otimes M_\psi(j).$$
We can also define the trace of an operator by $\trace(O):=(m_{\trace(O)},L_{\trace(O)},M_{\trace(O)},R_{\trace(O)})$ where $m_{\trace(O)}=m_O$, $L_{\trace (O)}=L_O$, $R_{\trace(O)}=R_O$ and $M_{\trace(O)}=\sum_{i\in\B} M_O(i,i)$.
Finally, we note that this formalism also gives us a way to construct density operators, i.e. for a pure state $\ket{\psi}=(m_\psi,L_\psi,M_\psi,R_\psi)$ we have that $\Psi:=\ketbra{\psi}{\psi}$ is given by $\Psi=\left(m_\Psi,L_\Psi,M_\Psi,R_\Psi\right)$ where $m_\Psi=m_\psi^2$, $L_\Psi = L_\psi \otimes L_\psi^*$,  $R_\Psi = R_\psi \otimes R_\psi^*$ and $M_\Psi(i,j) = M_\psi(i)\otimes M^*_\psi(j)$.  All of the operations discussed in this paragraph are consistent with applying the analogous operations in $\scalespace$ and $\scalespaceops$ after lifting states and operators to these space using $G$, as can be seen by writing down the sums and regrouping terms analogous to that done in Eq. \ref{eq:proof-operation-lifting}.

\section{Properties}
\label{sec:properties}

In this section we shall discuss several properties of $\mpsspace$.

\subsection{Translational invariance}

Given $\vs\in\scalespace$ and $\vz\in\basis$, let $\left[\vz^{\rightrightarrows k}\right]_i := \vz_{i-k}$, that is, the vector $\vz$ shifted $k$ sites to the right.  Observe that for all $i,k\in\Z$ we have that,
\begin{equation*}
\begin{aligned}
F(\vs)\paren{\vz^{\rightrightarrows k},i+k}
    &= n\mapsto L\cdot \paren{\prod_{j=0}^{n-1} M\paren{\left[\vz^{\rightrightarrows k}\right]_{i+j+k}}}\cdot R, \\
    &= n\mapsto L\cdot \paren{\prod_{j=0}^{n-1} M\paren{\vz_{i+j}}}\cdot R, \\
    &= F(\vs)\paren{\vz,i}.
\end{aligned}
\end{equation*}
The above implies that states in $\scalespace$ are translationally invariant because they only care about what they are seeing in the basis vector $\vz$, and ignore the absolute location.

\subsection{Transient behaviors}
\label{subsec:transient}

Let $\ket{\psi}\in\mpsspace$ be given by $m=1$, $L_1=1$, $R_1=1$, and $M_{11}(b)=\delta_{b0}$;  note that $G(\ket{\psi})=\prod_{i\in\Z}\ket{0}^i$.  Now consider the basis vector
$$\vz_i := \begin{cases}
1 & i=5 \\
0 & \text{otherwise}.
\end{cases}$$
Informally, we would expect the weight of the basis vector $\vz$ to be zero in the state $(F\circ G)(\ket{\psi})$ as the basis vector is 1 at position 5, which makes it orthogonal to $0^\Z$.  Despite this, we have that
$$F(\psi)(\vz,i)= \paren{n\mapsto \begin{cases}
0 & i \le 5 < i+n \\
1 & \text{otherwise}
\end{cases}} \ne 0.$$
The reason for this is that the finite-length slice of the basis vector which is used to calculate the coefficient only takes into account the behavior inside the slice, so when the slice does not include the mismatched site, the coefficient is non-zero.  Note, however, that for all $i\le 5$ and $n>5-i$ we have that $(F\circ G)(\ket{\psi})(\vz,i)=0$ --- that is, although $(F\circ G)(\ket{\psi})(\vz,\cdot)$ has a \emph{transient} behavior that is non-zero, the \emph{long-term} behavior is zero.  To be more precise, we say that the long-term behavior for the weight of a basis vector $\vz$ is zero if there exists $i',j'\in Z$ such that $i'\le j'$ and $(F\circ G)(\ket{\psi})(\vz,i)(n)=0$ for all $i\le i'$ and $n>j'-i$;  in particular, for the example we just discussed we have that $i'=j'=5$.  Conversely, the long-term behavior is non-zero when there exists an infinite sequence of pairs $\{i_k,j_k\}_{k\in\Z}$ such that for every $k\in\Z$ we have that $i_{k+1}<i_k < j_k <j_{k+1}$ and $(F\circ G)(\ket{\psi})(\vz,i_k)(j_k-i_k)\ne 0.$

The existence of zero long-term behaviors implies that there are states which are ``essentially'' orthogonal but have transient behaviors which mask this fact.  As the inner product can be computed exactly as a function, one can address this by declaring that two states with inner-product $f\in\N\to\C$ are orthogonal if and only if there exists $n$ such that for all $n'\ge n$, $f(n')=0$.

\subsection{Limiting behaviors}

One is often in the position of being primarily interested in the limiting behavior of a system.  Given a function in $\subring$, the limit as $n\to\infty$ is given by the dominant exponential term (or zero if there are no exponential terms) --- i.e.,
$$\lim_{n\to\infty} \sum_{k\in K}\lambda_k^n \cdot \text{poly}_k(n) + \sum_{l\in L}c_l \delta_{nl} = \lambda_{\max}\cdot\text{poly}_{\max}(n)$$
where $\lambda_{\max}$ is the eigenvalue with the largest magnitude (assuming it is not zero, in which case the right hand side would be a sum of delta functions).

Fortunately, computing the limiting form of the behavior is much less expensive than computing the exact form.  We start by recalling that the behavior of $\braket{\psi}{\psi}$ is given by $n\mapsto L'\cdot (M')^n\cdot R'$ where $L'=L^*\otimes L$, $R'=R^*\otimes R$, and $M'=\sum_{i\in\B} M^*(i)\otimes M(i)$.  Let $\lambda$ be the maximum eigenvalue of $M'$;  in the limit where $n\to\infty$, all of the Jordan blocks with eigenvalues less than $\lambda$ will effectively vanish as they become infinitely small relative to the Jordan blocks with eigenvalue $\lambda$.  Thus, it suffices to compute the dominant eigenvalue $\lambda$ and its associated eigenvectors, $\{v_i\}_{i\in I}$ (where $I$ is some index set).

In general $M'$ will not be diagonalizable, so the next step is to compute the Jordan blocks for each eigenvector;  this is done by computing the generalized eigenvectors $g_{ik}$ where $g_{i0}=v_i$ and $(M'-\lambda I)g_{ik}=g_{i(k-1)}$, obtaining vectors that satisfy $(M'-\lambda I)^{k+1} g_{ik}=0$.  At this point one has two pieces of information:  the size of each of the Jordan blocks, which is sufficient for us to compute the result of raising each to the power $n$ by using equation \eqref{jordanfunction}, and unitary matrices from the generalized eigenvectors such that $\lim_{n\to\infty}(M')^n =UJ^mU^{\dagger}$, where $J$ is in Jordan normal form with all of the blocks associated with $\lambda$.  This is sufficient for us to compute the limiting behavior of $\braket{\psi}{\psi}$ in closed form;  an analogous procedure can be used for computing the limiting behavior of $\left<\psi\left|O\right|\psi\right>$.

We finally note that if $O$ is in lower-triangular form, then it is possible to compute the limiting behavior without having to use any kind of eigensolver, but rather by solving a set of recurrence relations to get the fixed point;  see \cite{Michel2010}.

\subsection{Expectation normalization}

For a state $\ket{\psi}\in\mpsspace$, the normalization is given by $\braket{\psi}{\psi}\in\subring$, which, recall, is equivalent to the set of functions taking the form \eqref{expolyform};  as most elements in $\subring$ do not have a reciprocal, this means that $\ket{\psi}$ will not in general be normalizable, the exception being if $\braket{\psi}{\psi}=n\mapsto\lambda^n$, in which case we can let $\big|\tilde\psi\big>:=(n\mapsto\lambda^{-n/2})\ket{\psi}$ so that $\big<\tilde\psi\big|\tilde\psi\big>=n\mapsto 1$.

Because we cannot assume that states are normalized, expectation values must always take the form
$$\frac{\bra{\psi}O\ket{\psi}}{\braket{\psi}{\psi}},$$
which works because our space is the ring of functions $\newfield$, which, recall, has the property that quotients are fully defined as long as for all $n\in\N$, either $\braket{\psi}{\psi}(n)\ne 0$ or $\bra{\psi}O\ket{\psi}(n)= 0$, and if the latter condition holds, then the quotient is defined to be 0 for that value of $n$.

\section{Examples}
\label{sec:examples}

In this section we shall examine some examples of states in $\mpsspace$ and operators in $\mpsops$ and show how their inner products have the desired properties discussed in Section \ref{sec:motivation}.   We shall adopt the convention that
$$\begin{bmatrix}a \\ b\end{bmatrix}\otimes \begin{bmatrix}c \\ d\end{bmatrix} = \begin{bmatrix}ac \\ ad \\ bc \\ bd \end{bmatrix}, \quad\text{and similarly,}\quad
\begin{bmatrix}
a & b \\
c & d
\end{bmatrix}
\otimes
\begin{bmatrix}
w & x \\
y & z
\end{bmatrix}
=
\begin{bmatrix}
aw & ax & bw & bx \\
ay & az & by & bz \\
cw & cx & dw & dx \\
cy & cz & dy & dz \\
\end{bmatrix}.
$$

\subsection{Cat state}

Recall that the cat state is (informally) given by
$$\ket{\text{cat}} = \ket{0}^{\otimes\,\Z} + \ket{1}^{\otimes\,\Z}.$$

To have our state live in the space $\truespace$, we need to express this state as a function from $\basis\times\Z$ to $\newfield$.  The idea behind the following construction is that, for all $\vz\in\basis$, $i\in\Z$ and $n\in\N$, if the slice $\vz_{i:i+n}$ is equal to either $0^n$ or $1^n$ ($\in\B^{\otimes\,n}$), then it is mapped to 1;  otherwise, it is mapped to 0.  The following construction has exactly this property:
$$\ket{\text{cat}}(\vz,i) := n \mapsto
\begin{cases}
1 & \vz_{i:i+n} = 0^n\\
1 & \vz_{i:i+n} = 1^n \\
0 & \text{otherwise}.
\end{cases}$$

This is equivalent to $(F\circ G)(\ket{\psi})$ where $\ket{\psi}\in\mpsspace$ is given by
$$L:=\begin{bmatrix}1 & 1\end{bmatrix}, \quad R := \begin{bmatrix}1 \\ 1\end{bmatrix}, \quad M(0):= \begin{bmatrix}1 & 0 \\ 0 & 0\end{bmatrix}, \quad M(1):=\begin{bmatrix}0 & 0\\ 0 & 1\end{bmatrix}.$$

The norm, $\braket{\psi}{\psi}$, is given by $n\mapsto L_N\cdot M_N^n\cdot R_N$ where
$$L_N=L^*\otimes L =
\begin{bmatrix}
1 & 1 & 1 & 1
\end{bmatrix},
\quad R_N=R^*\otimes R =
\begin{bmatrix}
1 \\
1 \\
1 \\
1
\end{bmatrix},$$
$$M_N = \sum_{i\in\B}M^*(i)\otimes M(i) =
\begin{bmatrix}
1 & 0 & 0 & 0 \\
0 & 0 & 0 & 0 \\
0 & 0 & 0 & 0 \\
0 & 0 & 0 & 1 \\
\end{bmatrix}.
$$

It is straightforward to show based on this that $\braket{\psi}{\psi}=n\mapsto 2$;  if we wished, we could make the normalization 1 by multiplying the state by $n\mapsto 1/\sqrt{2}$.

\subsection{W state}

Recall that the W-state is given (informally) by$$\left\|\sum_{j=1}^n \bigotimes_{k=1}^n \begin{cases} \ket{1} & j=k \\ \ket{0} & \text{otherwise} \end{cases} \right\| = n.$$
Again, to have our state live in the space $\truespace$ we need to express this state as a function from $\basis$ to $\newfield$.  The idea behind the following construction is that for all $\vz\in\basis$, $i\in\Z$ and $n\in\N$, if the slice $\vz_{i:i+n}$ `looks like' a W-state then it is mapped to 1 and otherwise to 0.  That is,
$$\ket{W}(\vz,i)=n\mapsto
\begin{cases}
    1 & \vz_{i:i+n} \in\{0^{\,j}\,1\,0^{\,k}:j+1+k=n\}\subset\B^{\otimes\,n} \\
    0 & \text{otherwise}.
\end{cases}$$

This state is equal to $(F\circ G)(\ket{\psi})$ where $\ket{\psi}\in\mpsspace$ is given by
$$L=\begin{bmatrix}1 & 0\end{bmatrix},\quad R=\begin{bmatrix}0 \\ 1\end{bmatrix},\quad M(0) = \begin{bmatrix}1 & 0 \\ 0 & 1\end{bmatrix}, \quad M(1)=\begin{bmatrix}0 & 1\\ 0 & 0\end{bmatrix}.$$

The norm, $\braket{\psi}{\psi}$, is given by $n\mapsto L_N\cdot M_N^n\cdot R_N$ where
$$L_N=L^*\otimes L =
\begin{bmatrix}
1 & 0 & 0 & 0
\end{bmatrix},
\quad R_N=R^*\otimes R =
\begin{bmatrix}
0 \\
0 \\
0 \\
1
\end{bmatrix},$$
$$M_N = \sum_{i\in\B}M^*(i)\otimes M(i) =
\begin{bmatrix}
1 & 0 & 0 & 1 \\
0 & 1 & 0 & 0 \\
0 & 0 & 1 & 0 \\
0 & 0 & 0 & 1 \\
\end{bmatrix}.
$$

By swapping the second and fourth basis vectors --- i.e., using the basis transformation $U_{ij}=\delta_{(ij),(11)} + \delta_{(ij),(24)} + \delta_{(ij),(33)} + \delta_{(ij),(42)}$ --- we see that
$$\braket{\psi}{\psi} = n\mapsto \begin{bmatrix} 1 & 0 \end{bmatrix} \cdot \begin{bmatrix}
1 & 1 \\ 0 & 1
\end{bmatrix}^n \cdot \begin{bmatrix}0 \\ 1 \end{bmatrix} = n,$$
which matches the expected norm for the W state in Section \ref{sec:motivation}.

\subsection{Magnetic field}

Let $O\in\mpsops$ be the external magnetic field operator pointing against the z direction, which, recall, is given (informally) by $\sum_{k\in\Z} \sigma_{Z}^k$.

This has the same structure as the W-state, so its representation is similarly given by
$$L=\begin{bmatrix}1 & 0\end{bmatrix},\quad R=\begin{bmatrix}0 \\ 1\end{bmatrix},\quad 
M(i,j) = \begin{bmatrix} I_{ij} & (\sigma_Z)_{ij} \\ 0 & I_{ij} \end{bmatrix},$$
where $I$ is the identity matrix and $\sigma_Z$ is the Pauli Z spin matrix.

We now consider the expectation value of this operator on some of the previously considered states.

\subsubsection{Expectation of cat state}

The \emph{unnormalized} expectation of a cat state in a magnetic field is given by $L\cdot M^n\cdot R$ where
$$L = L_{\text{cat}}^*\otimes L_O\otimes L_{\text{cat}} = \begin{bmatrix}1 & 1\end{bmatrix}\otimes\begin{bmatrix}1&0\end{bmatrix}\otimes\begin{bmatrix}1 & 1\end{bmatrix}=\begin{bmatrix}1&1&0&0&1&1&0&0\end{bmatrix},$$
$$R = R_{\text{cat}}^*\otimes R_O\otimes R_{\text{cat}} = \begin{bmatrix}1 \\ 1\end{bmatrix}\otimes\begin{bmatrix}0\\1\end{bmatrix}\otimes\begin{bmatrix}1 \\ 1\end{bmatrix}=\begin{bmatrix}0&0&1&1&0&0&1&1\end{bmatrix}^{\text{T}},$$
$$M =\sum_{i,j=1}^2 M_{\text{cat}}^*(i)\otimes M_O(i,j)\otimes M_{\text{cat}}(j)=\begin{bmatrix}
1 & 0 & 1 & 0 & 0 & 0 & 0 & 0 \\
0 & 0 & 0 & 0 & 0 & 0 & 0 & 0 \\
0 & 0 & 1 & 0 & 0 & 0 & 0 & 0 \\
0 & 0 & 0 & 0 & 0 & 0 & 0 & 0 \\
0 & 0 & 0 & 0 & 0 & 0 & 0 & 0 \\
0 & 0 & 0 & 0 & 0 & 1 & 0 & -1 \\
0 & 0 & 0 & 0 & 0 & 0 & 0 & 0 \\
0 & 0 & 0 & 0 & 0 & 0 & 0 & 1 \\
\end{bmatrix}.
$$
We see from inspecting $M$ that rows and columns 2, 4, 5, and 7 are zeros, and hence can be eliminated from the system, so that $L\cdot M^n\cdot R \equiv L'\cdot (M')^n\cdot R'$ where
$$
L' =\begin{bmatrix}1&0&1&0\end{bmatrix},\quad
R'=\begin{bmatrix}0&1&0&1\end{bmatrix}^{\text{T}},\quad
M' =\begin{bmatrix}
1 & 1 & 0 & 0 \\
0 & 1 & 0 & 0 \\
0 & 0 &  1 & -1 \\
0 & 0 & 0 & 1 \\
\end{bmatrix}.
$$
We observe that this matrix is equal to the direct sum of two Jordan blocks;  if we break the linear system up into a direct sum of these blocks and their corresponding boundaries, then we observe that the expectation is given by the sum of the linear functions $n\mapsto +n$ and $n\mapsto -n$ which equals zero.  This is exactly what we would expect from such a system, as the component of the cat state which points up has an energy that is equal in magnitude but opposite in sign to the component that points down so that they cancel, resulting in a net energy of zero.

\subsubsection{Expectation of W state}

The \emph{unnormalized} expectation of a W state in a magnetic field is given by $L\cdot M^n\cdot R$ where
$$L = L_{\text{W}}^*\otimes L_O\otimes L_{\text{W}} = \begin{bmatrix}1 & 0\end{bmatrix}\otimes\begin{bmatrix}1&0\end{bmatrix}\otimes\begin{bmatrix}1 & 0\end{bmatrix}=\begin{bmatrix}1&0&0&0&0&0&0&0\end{bmatrix},$$
$$R = R_{\text{W}}^*\otimes R_O\otimes R_{\text{W}} = \begin{bmatrix}0 \\ 1\end{bmatrix}\otimes\begin{bmatrix}0\\1\end{bmatrix}\otimes\begin{bmatrix}0 \\ 1\end{bmatrix}=\begin{bmatrix}0&0&0&0&0&0&0&1\end{bmatrix}^{\text{T}},$$
$$
\begin{aligned}
M
=&\sum_{i,j\in\B}M_{\text{W}}*(i)\otimes M_O(i,j)\otimes M_{\text{W}}(j)\\
=&\begin{bmatrix}
1 & 0 & 1 & 0 & 0 & 1 & 0 & -1 \\
0 & 1 & 0 & 1 & 0 & 0 & 0 & 0 \\
0 & 0 & 1 & 0 & 0 & 0 & 0 & 1 \\
0 & 0 & 0 & 1 & 0 & 0 & 0 & 0 \\
0 & 0 & 0 & 0 & 1 & 0 & 1 & 0 \\
0 & 0 & 0 & 0 & 0 & 1 & 0 & 1 \\
0 & 0 & 0 & 0 & 0 & 0 & 1 & 0 \\
0 & 0 & 0 & 0 & 0 & 0 & 0 & 1 \\
\end{bmatrix}\\
\equiv
U^\dagger
&\begin{bmatrix}
1 & 1 & 1 & -1 & 0 & 0 & 0 & 0\\
0 & 1 & 0 & 1 & 0 & 0 & 0 & 0\\
0 & 0 & 1 & 1 & 0 & 0 & 0 & 0\\
0 & 0 & 0 & 1 & 0 & 0 & 0 & 0\\
0 & 0 & 0 & 0 & 1 & 1 & 0 & 0\\
0 & 0 & 0 & 0 & 0 & 1 & 0 & 0\\
0 & 0 & 0 & 0 & 0 & 0 & 1 & 0\\
0 & 0 & 0 & 0 & 0 & 0 & 0 & 1\\
\end{bmatrix}
U,
\end{aligned}
$$
where $U$ is a permutation matrix that moves rows and columns numbered 1, 3, 6, and 8 to the beginning, and the remaining rows and columns to the end;  because the boundary vectors only act non-trivially on the first and last (unpermuted) row and column, we conclude that only the upper-left corner of the permuted matrix is relevant, and so we can express the expectation in the form $L'\cdot (M')^n\cdot R'$ where $L'=\begin{bmatrix}1 & 0 & 0 & 0\end{bmatrix}, R'=\begin{bmatrix}0  & 0 & 0 & 1\end{bmatrix}^{\text{T}}$, and
$M'=\begin{bmatrix}
1 & 1 & 1 & -1 \\
0 & 1 & 0 & 1 \\
0 & 0 & 1 & 1 \\
0 & 0 & 0 & 1
\end{bmatrix}.$
We shall analyze the expectation by letting $\begin{bmatrix}a_n & b_n & c_n & d_n\end{bmatrix}^{\text{T}} := (M')^n\cdot R'$ and analyzing the sequences $a_n$ through $d_n$.  First, we observe that the last row is zero except for the last column, and so $d_n=1$.  Next, we observe that the last two rows and columns are self-contained and the submatrix on this space is equal to the Jacobian matrix for the linear function $n\mapsto n$, and thus $c_n=n$;  with similar reasoning we likewise conclude that $b_n=n$.  Finally, from the first row of $M'$ we know that $a_{n+1}-a_n=b_n+c_n-d_n=2n-1$, and since  $a_n=\sum_{i=1}^n (a_i-a_{i-1})$ (with $a_0=0$), we conclude that $a_n=\sum_{i=1}^n \paren{2(i-1) - 1} = n(n-1)-n=(n-1)^2-1$.

To obtain the \emph{normalized} expectation, we recall that the normalization of the $W$ state is given by $n\mapsto n$, so we therefore have that 
$$\left<\text{W}\right|O\left|\text{W}\right> = n\mapsto \frac{(n-1)^2-1}{n}=n+O(1).$$
(Note that the numerator equals zero when the denominator is zero, so that the right-hand-side is well-defined using our extended rule for division.)  In the large $n$ limit, the expectation grows linearly with the number of sites in the system, as expected.

\section{Relationship to similar work}
\label{sec:other-work}

In this section we discuss other work that is related, in that it deals with formalisms for working with the states of infinitely large systems.

\subsection{Infinite direct product construction}

In 1939, von Neumann built a formalism for working with infinite tensor products of Hilbert spaces that was based on the premise that the essential property of a tensor product of spaces is that each element in the space that was created using a tensor product should have its norm and inner product be equal to the product of, respectively, the norms and the inner products of each of the factors in the tensor product forming the element \cite{Neumann1939}.  To put this more precisely:  let $I$ be an index set (which may be uncountable), $\{\mathcal{H}_i\}_{i\in I}$ a (possibly uncountable) family of Hilbert spaces, and $\bigotimes_{i\in I} \mathcal{H}_i$ the infinite (von Neumann) tensor product of the family of Hilbert spaces.  Then, given two states $\bigotimes_{i\in I} f_i$ and $\bigotimes_{\in I} g_i$, the norm and inner product should be defined by, respectively, $\|\bigotimes_{i\in I} f_i\|:=\prod_{i\in I} \|f_i\|$ and $\paren{\bigotimes_{i\in I} f_i}\cdot\paren{\bigotimes_{i\in I} g_i}:=\prod_{i\in I} (f_i\cdot g_i)$.  Obviously these definitions only make sense if the products are convergent --- or, more precisely, \emph{quasi-convergent}\footnote{A product is quasi-convergent when the product of the absolute values converges;  if the phases do not converge, then the value of the product is defined to be zero.} --- so von Neumann \emph{starts} with the set of all tensor products that have quasi-convergent norms and then defines his infinite tensor product to be equal to the set of all finitary sums of these elements plus their limit points, the latter ensuring that the space is complete.

The construction described above is, in a sense, the unique way to construct infinite tensor product spaces because of a theorem (Theorem IV on p.33 of \cite{Neumann1939}) that shows that any complete Hilbert space that
\begin{enumerate}
\item includes an element corresponding to $\bigotimes_{i\in I} f_i$ for every quasi-convergent (in the sense of the norm) $\{f_i\}_{i\in I}$;
\item has the inner product of any two elements in item 1 be given by $\paren{\bigotimes_{i\in I} f_i}\cdot\paren{\bigotimes_{i\in I} g_i}:=\prod_{i\in I} (f_i\cdot g_i)$; and
\item has the full space be dense in the set of all finite sums of the elements in item 1
\end{enumerate}
is isomorphic to the construction in the previous paragraph.

von Neumann likewise constructed a set of operators acting on such spaces by starting with the set of operators acting non-trivially on exactly one of the Hilbert spaces and then taking the set of all operators generated by all finite sums and products of operators in this (starting) space as well as their limit points to obtain the full set of operators of interest acting on $\bigotimes_{i\in I} \mathcal{H}_i$.

The construction of these sets of infinite states and their infinite operators is incredibly elegant and very general, but it lacks the ability to represent the W-state or the external magnetic field operator, as both require a sum over an infinite number of terms --- in the case of the W state, terms such that one site points up and the rest point down, and in the case of the external magnetic field, terms such that the operator at one site is non-trivial and those at the rest are trivial.  Not only do these two constructions require an infinite number of terms, but the sums cannot be expressed as the limit of a sequence of partial sums because for any finite partial sum, the distance between the partial sum and the full sum is infinitely large (and therefore in particular for any $\epsilon>0$ there exists no subset of terms such that the difference between the partial sum and the full sum is below $\epsilon$).

On the other hand, infinite tensor product states are not required to be translationally invariant or even to exist on a countably infinite lattice;  thus, there are numerous examples of infinite tensor product states that do not exist in the formalism described in this paper.

\subsection{Finitely correlated states}

Because finitely correlated states are built on top of the (infinite) inductive limit of $C^*$-algebras, it is worth taking a moment to briefly review this formalism.  Unlike the formalisms considered so far in this paper, $C^*$-algebras are a construction of \emph{operators} acting on a Hilbert space rather than states living in a Hilbert space;  states (which by default are \emph{mixed} rather than pure) are then defined as \emph{functionals} from this operator space to complex numbers \cite{Bratteli1987}.  The use of an \emph{inductive} limit to define the $C^*$-algebra for the infinite chain precludes the existence of divergences because every element in the algebra of the infinite chain must exist in the algebra of some finite subset of the chain, which means it can only act non-trivially on a finite number of sites.

Within this context, a finitely correlated state (specialized to the $C^*$-algebra of operators over $\C^2\,\,$\footnote{The original definition is more general, with the physical space of operators being a $C^*$-algebra, the auxiliary space a general linear space, and the left boundary condition an element from the dual space of the auxiliary space.}) is given by a tuple $(\E,\rho,e)$ where $\rho,e\in\C^m$ and $\E\in\linearof(\C^2)\to\linearof(\C^m)$ is a tensor that maps linear operators in $\C^2$ to the ``auxiliary algebra'' of linear operators acting on $\C^m$;  furthermore, the components of this tuple must have the property that $\E(I)\cdot e=e$ and $\rho\cdot \E(I)=\rho\,\,$\footnote{Here we use a slightly different notation from that of Fannes et. al to better match the notation used in this paper;  in the original paper the element from the $\C^2$ was denoted by a subscript and the result was used as a map, so that these two properties were given as $\E_I(e)=e$ and $\rho\circ\E_I=\rho$.  We avoided using this notation in this paper because we have restricted ourselves to finite-dimensional spaces (mostly so that we can always compute the Jordan Normal form) and so there was nothing to be gained by using function composition notation rather than dot product notation.}.  The expectation of a local operator $\bigotimes_{i=1}^N O_i$ with $O\in\linearof(C^2)$ is given by $\left<O_1\otimes\cdots\otimes O_N\right>=\rho\cdot \E(O_1)\cdot\E(O_2)\cdots\E(O_N)\cdot e$.

Every finitely correlated state generated by $C:=(\E,\rho,e)$ is equivalent to an infinite density matrix $D:=(m,L,R,M)$ such that $m:=\dim \rho$, $L := \rho$, $R := e$, and $M(j,k):=\E(\ketbra{j}{k})$.  To see that they are equivalent, let $O$ be a local operator acting non-trivially on $N$ adjacent sites; then there exists an infinite matrix product operator $(m_O,L_O,M_O,R_O)$ such that $m_O:=N+1$, $L_i := \delta_{i1}$, $R_i := \delta_{(N+1)i}$, and
$$
M(i,j) := \begin{bmatrix}
I_{ij} & (O_1)_{ij} & 0 & \dots & 0\\
0 & 0 & (O_2)_{ij} &  \ddots & \vdots\\
\vdots &  & \ddots & \ddots & 0 \\
\vdots  &  & & 0 & (O_N)_{ij} \\
0 & \dots &\dots& 0 & I_{ij} \\
\end{bmatrix}.$$
Note that $G(O)=\prod_{n\in\N} T_n\in\scalespaceops$ where
$$
T_n =
\begin{cases}
0 & n < N \\
\sum_{i=0}^{n-N} I^{\otimes\,i}\otimes O_1\otimes O_2\cdots \otimes O_N \otimes I^{\otimes\,(n-N-i)} & \text{otherwise}.
\end{cases}
$$
Based on this we see that $$\left<O\right>_D=\trace(O\cdot D)=n\mapsto \begin{cases} 0 & n<N \\ (n-N+1)\left<O\right>_C & \text{otherwise},\end{cases}$$ and we do not need to divide by the normalization because for both states it is 1 as a consequence of the condition on $\E$.  Thus, for any local operator $O$ the expectation with respect to $C$ is the same as the expectation with respect to $D$.

Although every finitely correlated state is equivalent to an infinite density matrix in the formalism of this paper, the converse is not true because the condition on $\E$, that $\rho$ and $e$ be respectively left and right eigenvectors of $\E(I)$ with eigenvalue 1, will not always hold.  For example, consider the density matrix version of the W-state,
$$\sum_{i\in\Z}\bigotimes_{j\in\Z}\begin{cases}\ketbra{1}{1} & i = j \\ \ketbra{0}{0} & \text{otherwise}, \end{cases}$$
which you can think of as the fully mixed state where we know that exactly one of the qubits is in the $\ket{1}$ state but we do not know which.  This state corresponds to $(m,L,M,R)\in\mpsspace$ where $m=2$, $L=\delta_{i1}$, $R=\delta_{i2}$, and $M(i,j):=\begin{bmatrix}\delta_{(ij)(00)} & \delta_{(ij)(11)}\\ 0 & \delta_{(ij)(00)}\end{bmatrix}$.  If we were to represent this state as a finitely correlated state, we would have that $\rho:=L$, $e:=R$, and $\E(O):=\sum_{i,j\in\B} O_{ij}M(i,j)$;  in particular we have that $\E(I)=\begin{bmatrix}1&1\\0&1\end{bmatrix}$. There is only one right eigenvector of $\E(I)$, which is equal to $L$, and only one left eigenvector, which is is equal to $R$, and $L\cdot R=0$, so neither the left nor the right boundary vectors have non-zero overlap with an eigenvector (though they are equal respectively to the left and right generalized eigenvectors of order 2).  Thus, this state cannot be represented as a finitely correlated state.  \footnote{It is worth noting that this is not just a technicality because the whole reason for this property is that it allows one to essentially contract from infinity in both directions until one hits the non-trivial part of an operator;  if this property does not hold, then one no longer has a well-defined way to compute expectation values of operators using this formalism.}

The formalism of finitely correlated states also extends to pure states \cite{Fannes1994}.  
First, we define a subset of finitely correlated states called $C^*$-finitely correlated states where the auxiliary space is $\linearof(\C^m)$ for some $m$, $\rho$ and $e$ are positive-definite matrices, and $\E\in\linearof(\C^2)\to(\linearof \circ \linearof)(\C^m)$ is a completely positive map.  A $C^*$-finitely correlated state is then said to be purely generated if there exists an isometry $V:\C^m\mapsto\C^2\otimes\C^m$ such that $\E$ is given by $\E(O)_{(ij)(kl)}=\sum_{p,q=1}^2 V(b_k)_{pi}O_{pq}V(b_l)^*_{qj}$ where $(b_i)_j=\delta_{ij}$.  Because such states are a subset of finitely correlated states, they all correspond to infinite density matrices.  It is not the case, however, that they correspond to pure infinite matrix product states.  The reason for this is that the boundary conditions $\rho$ and $e$ are not required to be factorizable into a tensor product, i.e. there need not be $x,y\in\C^m$ such that $\rho=x\otimes y$.  We can, however, express such states as a sum of density matrices constructed from pure states as follows.  First, recall that $\rho$ and $e$ are positive-definite matrices (since we are talking specifically about $C^*$-finitely correlated states), which means that that $\rho=U_\rho\Lambda_\rho U_\rho^\dagger$ and $e=U_e\Lambda_e U_e^\dagger$.  Now for each $1\le i,j\le m$ we define $\ket{\psi_{ij}}:=(m,L_{ij},R_{ij},M)$ where $L_{ij}:=(U_\rho)_i\cdot\sqrt{(\Lambda_\rho)_{ii}}$, $R_{ij}:=(U_e)_j\cdot\sqrt{(\Lambda_e)_{jj}}$, and $M(i)_{jk}:=V(b_k)_{ij}$ (where again, $(b_i)_j=\delta_{ij}$).  We then have that the purely generated $C^*$-finitely correlated state is equivalent to $\sum_{i,j=1}^m\ketbra{\psi_{ij}}{\psi_{ij}}$.  In general, though, pure infinite matrix product states do not correspond with purely generated states because the matrix $M$ need not be an isometry, with the counter-example again being the W-state for which $M(i)=\begin{bmatrix}\delta_{i0}&\delta_{i1}\\0&\delta_{i0}\end{bmatrix}$;  to see why this is the case note that the tensor $V$ corresponding to $M$ is given by $V=\begin{bmatrix}\ket{0}&\ket{1}\\0&\ket{0}\end{bmatrix}$ and so we see that $V\cdot\begin{bmatrix}0\\1\end{bmatrix}=\begin{bmatrix}\ket{1}\\\ket{0}\end{bmatrix}$ which implies that $V$ is not an isometry because it can in general change the length of its input vector (in this case from 1 to $\sqrt{2}$).

Finally, we note that, although technically the formalism of finitely correlated states only defines expectation values for local operators, because every finitely correlated state is equivalent to an infinite density matrix, we can use this connection to define expectation values for all matrix product operators, thus employing the formalism of this paper to extend the set of operators for which expectations of finitely correlated states are defined.

\subsection{Infinite matrix product states with derived (post-hoc) boundaries}

When performing simulations with infinite matrix product states, the result often takes the form of an infinitely repeated middle tensor\footnote{For simplicity we will assume that there is only one such tensor unless mentioned otherwise, but in general there might be multiple such tensors depending on the canonical form used and the size of the unit cell.} that does not come with left and right boundaries (see \cite{Vidal2007}, \cite{Orus2008}, \cite{Crosswhite2008}, and \cite{McCulloch2008}).  Instead, the left and right boundaries are derived from the middle tensor by computing the dominant eigenvectors of the transfer matrix.  Put another way, we are given the (isometric) tensor $V$ for a purely generated finitely correlated state, and then $\rho$ and $e$ are constructed using the respective dominant left and right eigenvectors of the map $\E$;  if the dominant eigenvalue is not 1, then we can divide $V$ by the square root of the dominant eigenvalue to make it be 1.  Thus we see that all such infinite matrix product states are really just a form of purely generated finitely correlated state where the boundaries, $\rho$ and $e$, are completely dependent on the repeated tensor, $V$.

This connection does make a couple of assumptions, however:  first, it assumes that the result of the simulation is a tensor that is (or can equivalently be expressed as) an isometry, and second, it assumes that $\E$ only has a single dominant left and right eigenvector.  These assumptions are perfectly reasonable because they tend to be true for the set of tensors that are output by simulations, but they are very restrictive, and in particular do not hold for the W-state. \footnote{In particular, the canonicalization procedure in \cite{Orus2008} breaks down for the W-state in two places:  first, because the transfer matrix has multiple dominant eigenvectors (and one of them is a generalized eigenvector), and second, because reshaping any of these eigenvectors into a matrix results in an operator that is not full rank and so does not have an invertible square root.}

\section*{Conclusion}

In this paper we have presented a formalism for representing quantum states and operators for infinite systems that handles divergences by using maps instead of complex numbers for the space of coefficients.  Furthermore, we showed how this gives us additional power by allowing us to represent, manipulate, and analyze states and operators that are excluded by related formalisms.  This work can be used as a tool for studying characteristics of infinite matrix product states, and in particular their relation to constructions in the field of formal language theory.

\bibliographystyle{vancouver}
\bibliography{paper}

\end{document}